\documentclass[review,11pt]{elsarticle}

\biboptions{sort&compress}  %For Elsevier class

\usepackage[mathscr]{eucal}  %Sample space \[\mathscr{ABCDEFGHIJKLMNOPQRSTUVWXYZ}\]
\usepackage{graphicx,multirow} %metafile to eps converter
\usepackage[normal]{subfigure}
\usepackage{amssymb,amsmath} %\lessgtr </> threshold'la karsilastirma sembolu
\usepackage{accents}
\usepackage{mathtools}  %bmatrix*
\usepackage{booktabs}
\usepackage{cancel}

\usepackage{tikz}  %For circled numbers, \circled{1} \circled{2} \circled{3}
\newcommand*\circled[1]{\tikz[baseline=(char.base)]{
            \node[shape=circle,draw,inner sep=2pt] (char) {#1};}}

\usepackage[linesnumbered,ruled]{algorithm2e}

\newcommand{\be}{\begin{eqnarray}}
\newcommand{\ee}{\end{eqnarray}}
\newcommand{\nn}{\nonumber}
\newcommand{\mbf}[1]{\mbox{$\mathbf #1$}}    %Mathboldface
\newcommand{\mref}[1]{(\ref{#1})}

\newcommand{\vcase}[1]{case\,--\,\circled{#1}}
\newcommand{\BICN}{\mbox{$\textrm{BIC}_\textrm{N}$}}
\newcommand{\BICSNR}{\mbox{$\textrm{BIC}_{\textrm{SNR}}$}}

\newif\ifmyshow %% to be used with \usepackage[nomarkers]{endfloat}
\myshowfalse %% Enable this one if not

\journal{Signal Processing}

\begin{document}

\begin{frontmatter}
\title{Polynomial Order Selection for Savitzky–Golay Smoothers via N-fold Cross-Validation (extended version)}

\author[METU]{\c{C}a\u{g}atay Candan}
\ead{ccandan@gmail.com}

\address[METU]{Department of Electrical and Electronics Engineering, Bogazici University, Bebek, Istanbul, 34342, Turkey}

\begin{abstract}
Savitzky-Golay (SG) smoothers are noise suppressing filters operating on the principle of projecting noisy input onto the subspace of polynomials. A poorly selected polynomial order results in over- or under-smoothing which shows as either bias or excessive noise at the output. In this study, we apply the N-fold cross-validation  technique (also known as leave-one-out cross-validation) for model order selection and show that the inherent analytical structure of the SG filtering problem, mainly its minimum norm formulation, enables an efficient and effective order selection solution. More specifically, a novel connection between the total prediction error and SG-projection spaces is developed to reduce the implementation complexity of cross-validation method. The suggested solution compares favorably with the state-of-the-art Bayesian Information Criterion (BIC) rule in non-asymptotic signal-to-noise ratio (SNR) and sample size regimes. MATLAB codes reproducing the numerical results are provided.
\end{abstract}

\end{frontmatter}

\section{Introduction}
\label{SecIntro}
Savitzky-Golay (SG) filters are used for smoothing noisy input while preserving local polynomial characteristics such as linear, quadratic or higher order variations \cite{golay64,OrfaBOOK,ccandanSG}. SG filters are low-pass filters; however, unlike conventional filters, their spectral properties, such as cut-off frequency, pass-band ripple, side-lobe levels \cite{Schafer2011}, are secondary to the time-domain properties.

SG filters operate on the principle of i) projecting input onto a subspace of polynomials in the least squares sense, ii) mapping the projection result to a scalar output by evaluating the projection result (smoothing or interpolation application \cite{golay64}) or its derivative (differentiator application \cite{ccandanSG}) at a point, as discussed in \cite[Sec. 8.3.5]{OrfaBOOK} and   \cite{ccandanSG}. It is possible to express the cascade of these two operations as a finite impulse response (FIR), linear time-invariant (LTI) filter \cite{ccandanSG}. In many applications, the local characteristics of the signal are expected to be a polynomial function particularly in the instrumentation and measurement area \cite{SGappPLL2022,AdapSG2021}.
\ifmyshow \textcolor{red}{--- Figure 1 about here --- } \fi

\begin{figure}[h]
\centering
\includegraphics[height=10cm]{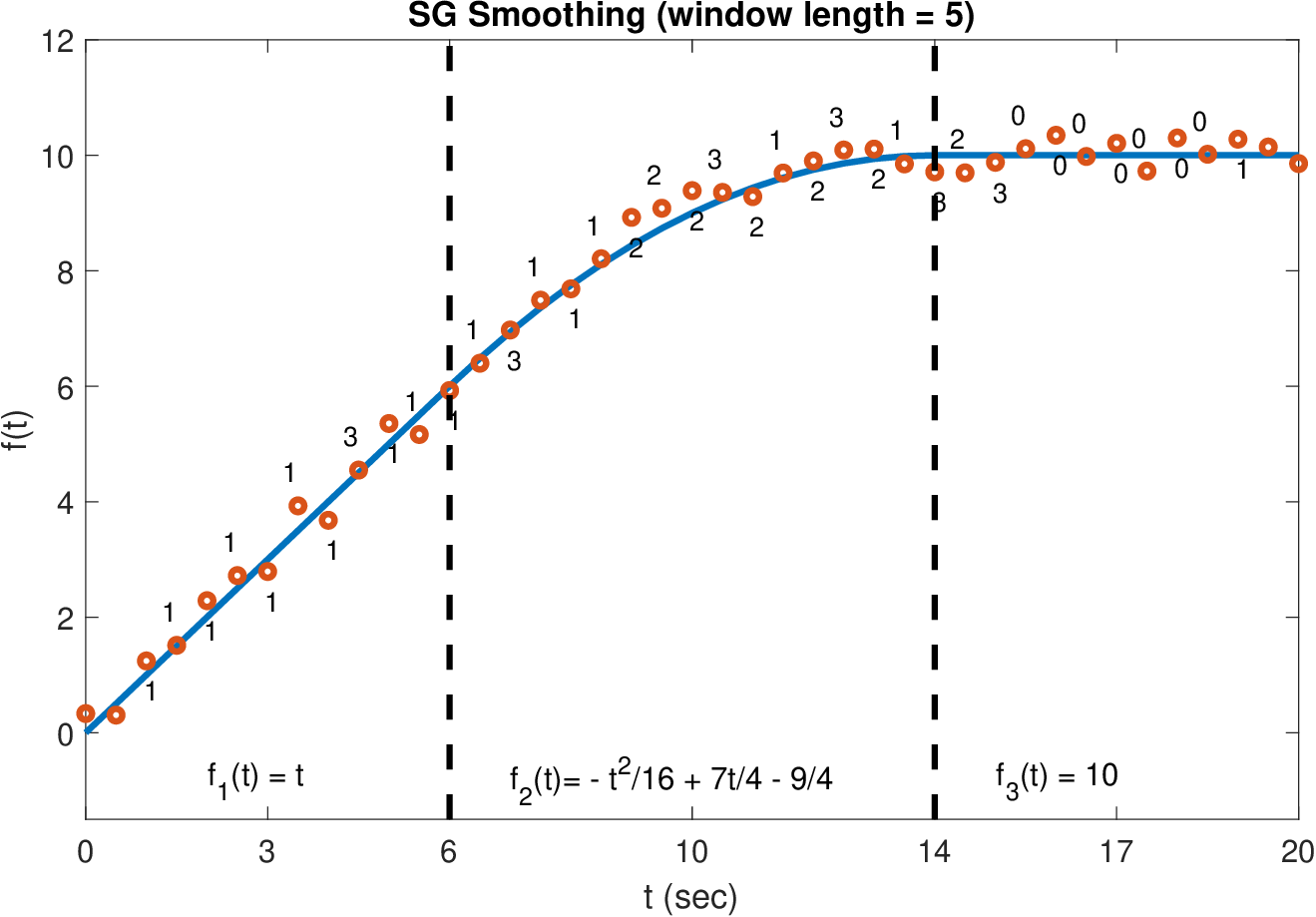}
\caption{An SG-smoothing example. The samples (circles) are off the piecewise-defined function due to noise. The numerical values above or  below of each sample are the selected polynomial order with the cross-validation technique.}
\label{fig1SG}
\end{figure}

Figure~\ref{fig1SG} illustrates a smooth, piecewise-defined polynomial function that is discretized with a sampling period of $T=0.5$ sec. The samples shown deviate from the function due to sampling noise. This function can be considered to represent the position of an object, say a satellite being tracked, which is at the origin at $t=0$ with an initial velocity of $1$ m/sec. The object maintains this velocity in the time interval of $t \in (0,6)$.  Over $t \in (6,14)$, the object decelerates  at a rate of $-0.125 \, \textrm{m}/\textrm{sec}^2$ and comes to a stop at $10$ meters from its initial position. The signal processing goal in relation to this kinematic experiment is the estimation of velocity, acceleration and other motion parameters from noisy position measurements in Figure~\ref{fig1SG}. The numerical values above or below each time sample show the best polynomial order fitted to that particular data point via the suggested N-fold cross validation method. Here, the maximum possible polynomial order is chosen as $3$. We observe that for $t \in (0,6)$, the best fit is the 1st degree except at one sample. In the 2nd segment with de-acceleration, the chosen order changes according to the effect of noise on local curvature; but it generally corresponds well to a 2nd degree trend.  In the final segment of $(14,20)$, the order is correctly predicted in the majority of samples.

SG filters have only two design parameters, the processing window length and the maximum polynomial degree. The processing window length, which is also the filter impulse response length, should be selected as long as possible to admit a reliable estimation of the assumed polynomial variation. In Figure~\ref{fig1SG}, the processing window length is $5$ samples, two neighboring samples on the immediate left and right of each target sample. In general, the choice of processing window length depends on the signal characteristics in relation with the signal bandwidth \cite{GStrang2003,WindowSelect2020,SG_IEEESP2013}.

Here we focus on selecting the best polynomial order at a fixed window length, that is the goal is to select the best polynomial order for each segment/sample of the data, as in Figure~\ref{fig1SG}. The key theoretical contribution is the development of an expression between the total prediction error, which is the cross-validation metric, and the projection spaces of SG-smoothing filters. This key relation not only explains the observed success of CV in this application and also reduces its implementation complexity significantly.

%\vspace*{-0.5cm}
\section{Background}
\label{SecBG}
\vspace*{-0.3cm}
\textbf{SG-smoothing Filters:} SG-smoothing filters can be designed as the solution of a minimum Euclidean norm problem \cite{ccandanSG}. As a concrete example, the SG smoother of length $5$ for the 2nd degree polynomials can be formulated as the minimum norm solution of
\be
\underbrace{
\left[
\begin{array}{ccccc}
 1 & 1  & 1 & 1 & 1 \\
 1 & 2  & 3 & 4 & 5 \\
1^2 & 2^2 & 3^2 & 4^2 & 5^2
\end{array}\right]}_{\mathbf{A}}
\left[
\begin{array}{c}
h_2 \\ h_1 \\ h_0 \\ h_{-1} \\ h_{-2}
\end{array}\right]
=
\left[
\begin{array}{c}
1 \\ 3 \\ 3^2
\end{array}\right].
\label{SGfilt}
\ee
If we denote the equation system in \mref{SGfilt} as $\mbf{A}\mathbf{h} = \mbf{b}$, the minimum Euclidean norm solution becomes $\mathbf{h}_{\textrm{MN}} = \mbf{A}^T(\mbf{A}\mbf{A}^T)^{-1}\mbf{b}$. Here $\mathbf{h}_{\textrm{MN}}$ is the impulse response of an FIR filter of length $5$, called SG-smoother, that projects the input onto the 3 dimensional sub-space spanned by the 2nd degree polynomials. In this case, the convolution operation $y_n = \sum_{k=-2}^2x_{n-k}h_k$ produces the smoothed output for the central sample $x[n]$ in the processing window of $\{x_{n-2},x_{n-1}, x_n, x_{n+1}, x_{n+2}\}$.

The rows of the $\mbf{A}$ matrix in \mref{SGfilt} are samples of the $0$th, $1$st and $2$nd order polynomials in $[1,5]$ with the unit sampling period. Observe that the 3 constraint equations in \mref{SGfilt} enforce the mapping of constant ($f_0(t) = 1$), linear ($f_1(t) = t$)  and quadratic ($f_2(t) = t^2$) function samples in $t \in [1,5]$ to the value of the central sample of the function at $t=3$. By linearity, an arbitrary 2nd degree polynomial $g(t) = \alpha_1 f_1(t) + \alpha_2 f_2(t) + \alpha_3 f_3(t)$  is mapped to $g(3)$ with the same transformation. Observe that any vector $\mbf{h}$ vector satisfying the equation system in \mref{SGfilt} establishes this mapping. What is unique about the minimum norm solution $\mathbf{h}_{\textrm{MN}}$ is that it is the only solution in the range space of $\mbf{A}^T$ (i.e., row-space of $\mathbf{A}$ matrix). Since the row space of $\mathbf{A}$ is the sub-space of 2nd degree polynomials, this implies that the vector $\mathbf{h}_{\textrm{MN}}$ is orthogonal to all 3rd or higher degree polynomials; thus, performing  the projection step in the two-step SG filtering procedure described in Section~\ref{SecIntro} \cite[Sec. 8.3.5]{OrfaBOOK}.
\ifmyshow \textcolor{red}{--- Table 1 about here ---} \fi

\begin{table}[h]
\centering
\caption{5-fold cross-validation example with $x_n = 7n^2 - 3$.}
\begin{tabular}{|c|ccccc|}
\hline
case \# & $x_{-2}$   & $x_{-1}$   & $x_0$      & $x_1$      & $x_2$      \\ \hline
\circled{1} &\color{red}{\textbf{?}} & 4          & -3         & 4          & 25         \\
\circled{2} &25         & \color{red}{\textbf{?}}  & -3         & 4          & 25         \\
\circled{3} &25         & 4          & \color{red}{\textbf{?}}  & 4          & 25         \\
\circled{4} &25         & 4          & -3         & \color{red}{\textbf{?}}  & 25         \\
\circled{5} & 25         & 4          & -3         & 4          & \color{red}{\textbf{?}}  \\ \hline
\end{tabular}
\label{SGvTable}
\end{table}

\textbf{Cross-Validation:} Cross-validation partitions the data into two sets, namely a training and a validation set. The goal is to select the hyperparameters of a model such as the number of clusters for the k-means algorithm or the polynomial degree for SG-smoothers, by comparing the prediction performance of different models trained (or optimized) on the training set and evaluated on the validation set \cite[Ch.13]{Barberbook}.

Continuing with the running example, let us assume that the samples in the processing window of $5$ samples are $x_{-2} = x_{2} = 25$, $x_{-1}= x_1 = 4$ and $x_0 = -3$, which corresponds to $x_n = 7n^2 - 3$ sequence. For $N$-fold cross-validation, the data of length $N$ (5 samples for this case) is partitioned $N$ times by setting aside a single sample as the validation set and using the remaining $N-1$ samples as the training set. The goal is to predict the validation sample by constructing a model from the training data. Table~\ref{SGvTable} shows five cases for the partitioning of $x_n$ with $5$-fold cross-validation. The samples with the question mark indicate the validation data to be predicted.

Considering the $0$th order model and \vcase{1}, we predict $x_{-2}$ by fitting a $0$th order model to $\{x_{-1},x_0,x_1,x_2\}$. In this case, the prediction is $\hat{x}_{-2} = 15/2$ and is formed by the arithmetic average of the training data. Since $x_{-2} = 25$, the prediction error is $ 25 - 15/2 = 35/2$. The prediction errors for all cases are computed similarly and the total squared prediction error, i.e., the sum of squared prediction errors for all cases, is recorded. This process is repeated for the 1st, 2nd and higher order models and the total squared prediction error for each model is noted. Once the process finalizes, the model with the smallest total squared prediction error is selected.

The cross-validation procedure does not require any additional parameters such as noise variance, signal power, etc. However, the process is highly repetitive and can be computationally intensive especially for time-series type data where a new model order is selected for each data point, as in Figure~\ref{fig1SG}.

\textbf{SG-Prediction Filter:} SG-prediction filters are needed to implement the cross-validation procedure. For illustration purposes, we focus on the SG-prediction with 2nd degree polynomials for \vcase{1} in Table~\ref{SGvTable}. The SG-predictor filter for this setting is the minimum norm solution of the following equation:
\be
\left[
\begin{array}{cccc}
1  & 1 & 1 & 1 \\
2 & 3 & 4 &5 \\
2^2 & 3^2 & 4^2 & 5^2
\end{array}\right]
\left[
\begin{array}{c}
h^{\textrm{p\circled{1}}}_{-1} \\
h^{\textrm{p\circled{1}}}_{-2} \\
h^{\textrm{p\circled{1}}}_{-3} \\
h^{\textrm{p\circled{1}}}_{-4}
\end{array}\right]
=
\left[
\begin{array}{c}
1 \\ 1 \\ 1^2
\end{array}\right].
\label{SGpredfilt}
\ee
If we compare the equations for the prediction and smoothing operations, we see that the first column of the matrix $\mbf{A}$ in the smoothing operation given in \mref{SGfilt} is missing on the left-hand side of \mref{SGpredfilt}; however, the same column appears on the right side of \mref{SGpredfilt}. It is worth noting that the constraint equations in \mref{SGpredfilt} map the samples of $x(t)=t^k$ collected at integer valued points in $[2,5]$ to the value of $x(1)$, which is the error-free prediction output.

The minimum norm solution of \mref{SGpredfilt} is $\mbf{h}^{\textrm{p\circled{1}}}_{\textrm{MN}} = \frac{1}{4}[9,\,\, -3,\,\, -5, \,\, 3]^T$ and the corresponding prediction for Case-\circled{1} is $\hat{x}_{-2} = (\mbf{h}^{\textrm{p\circled{1}}}_{\textrm{MN}})^T \times [x_{-1},\,\, x_0,\,\, x_{1},\,\, x_{2}] = 25$, which is identical to $x_{-2}$ value. This result is expected, since $x_n$ satisfies a quadratic relation ($x_n = 7n^2 - 3$) in this example.

%\vspace*{-0.5cm}
\section{SG Model Order Selection with N-fold Cross-Validation}
\label{SecMain}
We consider the design of an SG-smoothing filter with a processing window of length $N$ (also the filter impulse response length) projecting the input onto the $P+1$ dimensional subspace of $P$th degree polynomials. The constraint equations for the SG-smoother are:
\be
\underbrace{
\left[
\begin{array}{cccccccc}
 1 & 1  & \ldots &1  & 1 & \ldots & 1 & 1 \\
 1 & 2 & \ldots  &k-1& k & \ldots &  (N-1) & N \\
 \vdots & \vdots &\  & \vdots & \vdots & \  & \vdots & \vdots \\
1 & 2^P & \ldots &(k-1)^P& k^P & \ldots &  (N-1)^P & N^P
\end{array}\right]}_{\mathbf{A}}
\mbf{h}^{s}_k =
\underbrace{
\left[
\begin{array}{c}
1 \\ k \\ \vdots \\ k^P
\end{array}\right]}_{\mbf{c}_k}.
\label{SGfiltgen}
\ee
The equation system in \mref{SGfiltgen} has the vector $\mbf{c}_k$ on its right side. Here $\mbf{c}_k$ also refers to the $k$th column of the matrix $\mbf{A}$. The index $k$ indicates that the filter is constructed to smooth the $k$th sample in the processing window.  The minimum norm solution of \mref{SGfiltgen}, which yields the SG-smoother, is simply $\mbf{h^{s}_k} = \mbf{A}^T(\mbf{A}\mbf{A}^T)^{-1} \mbf{c}_k$. To smooth all $N$ samples in the processing window, from $k=1$ to $k=N$, we can construct $\mathbf{H}^s = [\mathbf{h}^s_1\,\, \mathbf{h}^s_2 \,\, \ldots \,\, \mathbf{h}^s_N] = \mbf{A}^T(\mbf{A}\mbf{A}^T)^{-1} \mbf{A}$ and the output of the smoother is then $(\mathbf{H}^s)^T \mbf{x}$. Here
$\mbf{x} = [ x_1\,\, x_2  \,\, \ldots \,\, x_N ]^T$ is the input data in the processing window.

SG-prediction filter for the prediction of the $k$th sample in the processing window can be similarly expressed as the minimum norm solution of the following system:
\be
\underbrace{
\left[
\begin{array}{cccccccc}
 1 & 1  & \ldots      & 1   & 0 & \ldots & 1 & 1 \\
 1 & 2  & \ldots      & k-1 & 0 & \ldots &  (N-1) & N \\
 \vdots & \vdots & \  & \vdots  & \vdots & \  & \vdots & \vdots \\
1 & 2^P & \ldots      & (k-1)^P  & 0 & \ldots &  (N-1)^P & N^P
\end{array}\right]}_{\mathbf{A_p}}
\mbf{h^{p}_k} =
\underbrace{
\left[
\begin{array}{c}
1 \\ k \\ \vdots \\ k^P
\end{array}\right]}_{\mbf{c}_k}.
\label{SGfiltpregen}
\ee
The equation systems in \mref{SGfiltgen} and \mref{SGfiltpregen} are nearly identical, differing only in the $k$th columns of the matrices $\mbf{A}$ and $\mbf{A}_p$. The matrix $\mbf{A_p}$ in \mref{SGfiltpregen} has an all zeros vector in its $k$th column, in contrast to matrix $\mbf{A}$ in \mref{SGfiltgen}. Note that the $k$th entry of the minimum norm solution to the equation system in \mref{SGfiltpregen} is identically $0$, that is $[\mbf{h^{p}_k}]_k = 0$. Hence, the $k$th column of $\mathbf{A_p}$ in \mref{SGfiltpregen} can be discarded and $\mbf{h^{p}_k}$  can be shortened without any harm. When this column is removed, we retrieve the special case of the SG-prediction filter in \mref{SGpredfilt} for $k=1, N=5, P=2$. In the following, we utilize the redundant $\mathbf{A}_p$ matrix with an all zeros column in the SG-predictor constraint equations to establish a connection with matrix $\mbf{A}$ in \mref{SGfiltgen}. Later, we capitalize on this connection to simply the cross-validation calculations.

\textbf{Prediction Error Calculation:} The prediction error for the $k$th sample in processing window is defined as $\epsilon_k^p \triangleq x_k - \hat{x}^p_k$ where $\hat{x}^p_k = (\mathbf{h}^{p}_k)^T \mbf{x}$ is the predicted value. The prediction error can also be expressed as $\epsilon_k^p = (\mathbf{e}_k^T - (\mathbf{h}^{p}_k)^T)\mbf{x}$ where $\mbf{e}_k =[0 \,\, 0 \,\, \ldots \,\, 1 \,\, \ldots \,\, 0]$ is the $k$th canonical basis vector in $N$-dimensional space, with all zero entries except a $1$ at the $k$th entry.

We note that the constraint equations for smoothing and prediction operations are linked with $\mbf{A}_p = \mbf{A} - \mbf{c}_k \mbf{e}_k^T$; since $\mathbf{A}_p$ in \mref{SGfiltpregen} differs from $\mbf{A}$ in \mref{SGfiltgen} only in the $k$th  column. If we substitute this equality in the expression for the minimum norm solution $\mbf{h}^{p}_k = \mathbf{A}_p^T(\mathbf{A}_p \mathbf{A}_p^T)^{-1}\mathbf{c}_k$, we get
\be
\mbf{h}^{p}_k & \stackrel{}{=} &
(\mbf{A} - \mbf{c}_k \mbf{e}_k^T)^T
\left(
(\mbf{A} - \mbf{c}_k \mbf{e}_k^T)
(\mbf{A} - \mbf{c}_k \mbf{e}_k^T)^T
\right)^{-1}\mathbf{c}_k \nn \\
& \stackrel{(a)}{=}  &
(\mbf{A} - \mbf{c}_k \mbf{e}_k^T)^T
\left(
\mbf{A}\mbf{A}^T - \mbf{c}_k \mbf{c}_k^T
\right)^{-1}\mathbf{c}_k \nn \\
& \stackrel{(b)}{=}  &
(\mbf{A} - \mbf{c}_k \mbf{e}_k^T)^T
\left(
(\mbf{A}\mbf{A}^T)^{-1} +
\frac{(\mbf{A}\mbf{A}^T)^{-1} \mathbf{c}_k \mathbf{c}_k^T (\mbf{A}\mbf{A}^T)^{-1} }
{1 - \mathbf{c}_k^T (\mbf{A}\mbf{A}^T)^{-1} \mathbf{c}_k}
\right) \mbf{c}_k
\nn \\
& \stackrel{(c)}{=}  &
\left( 1 +
\frac{\mathbf{c}_k^T (\mbf{A}\mbf{A}^T)^{-1} \mathbf{c}_k}
{1 - \mathbf{c}_k^T (\mbf{A}\mbf{A}^T)^{-1} \mathbf{c}_k}
\right) \mbf{A}^T(\mbf{A}\mbf{A}^T)^{-1} \mbf{c}_k
\nn \\
& & \quad \quad \quad  -
\left( \mathbf{c}_k^T (\mbf{A}\mbf{A}^T)^{-1} \mathbf{c}_k +
\frac{(\mathbf{c}_k^T (\mbf{A}\mbf{A}^T)^{-1} \mathbf{c}_k)^2}
{1 - \mathbf{c}_k^T (\mbf{A}\mbf{A}^T)^{-1} \mathbf{c}_k}
\right) \mbf{e}_k .
\label{hsgp}
\ee
In line-(a), the argument of matrix inverse operation is expanded and simplified by using $\mbf{c}_k = \mbf{A}\mbf{e_k}$. In line-(b), the matrix inversion lemma (Woodbury formula) is utilized \cite[Appendix A]{Barberbook}. In line-(c), the product is expanded and the result is observed to be a linear combination of two vectors, namely $\mbf{h^{s}_k} = \mbf{A}^T(\mbf{A}\mbf{A}^T)^{-1} \mbf{c}_k$ (SG-smoother for the $k$th sample)
and the canonical basis vector $\mbf{e}_k$. For some more simplifications, we note that
\be
\mathbf{c}_k^T (\mbf{A}\mbf{A}^T)^{-1} \mathbf{c}_k
= \mathbf{e}_k^T \mbf{A}^T (\mbf{A}\mbf{A}^T)^{-1} \mbf{A} \mathbf{e}_k
= \mathbf{e}_k^T \mbf{P_{A^T}} \mbf{e}_k = [\mathbf{P}_{A^T}]_{kk}.
\ee
Here $\mbf{P}_{A^T} = \mbf{A}^T (\mbf{A}\mbf{A}^T)^{-1} \mbf{A}$ is the projection matrix to the range space of $\mbf{A}^T$ (or to the row space of $\mbf{A}$) and $[\mathbf{P}_{A^T}]_{kk}$ is the $k$th row and $k$th column entry of this matrix. Given these, we can rewrite $\mbf{h}^{p}_k$ in \mref{hsgp} as follows:
\be
\mbf{h}^{p}_k =
\frac{1}{1 - [\mathbf{P}_{A^T}]_{kk}}\mbf{h}^{s}_k
-
\frac{[\mathbf{P}_{A^T}]_{kk}}{1 - [\mathbf{P}_{A^T}]_{k,k} }\mbf{e}_k.
\label{hsgp2}
\ee
After \mref{hsgp2}, the prediction error $\epsilon_k^p = x_k - \hat{x}^p_k  = (\mathbf{e}_k - \mathbf{h}^{p}_k)^T \mbf{x}$  simplifies to
\be
\epsilon_k^p = \frac{1}{1 - [\mathbf{P}_{A^T}]_{kk}}(\mbf{e}_k - \mbf{h}^{s}_k)^T \mbf{x}
=  \gamma_k (x_k - \hat{x}_k^s ) = \gamma_k \epsilon_k^s
\label{SGperr1}
\ee
where $\gamma_k = 1/(1 - [\mathbf{P}_{A^T}]_{kk})$, $\hat{x}_k^s = (\mbf{h}^{s}_k)^T \mbf{x}$ is the SG-smoother output and $\epsilon_k^s \triangleq x_k - \hat{x}_k^s$ is the residual error after smoothing. The residual error vector for all samples can then be written as
$
\boldsymbol{\epsilon}^s = [\epsilon^s_1 \,\, \epsilon^s_2 \,\, \ldots \,\, \epsilon^s_{N}]^T
 = (\mbf{I} - \mathbf{P}_{A^T}) \mbf{x} = \mathbf{P}_{A^T}^\perp \mbf{x}
$
where $ \mathbf{P}_{A^T}^\perp  = \mbf{I} - \mathbf{P}_{A^T}$ is the complementary projection matrix, that is the projector to the null space of $\mbf{A}$.

From the results obtained, the total squared prediction error, $T^p_\epsilon$, can be written as
\be
T^p_\epsilon = \sum_{k=1}^{N} (\epsilon_k^p)^2  = \sum_{k=1}^{N} \gamma_k^2 (\epsilon_k^s)^2.
\label{Tpe}
\ee
where $\gamma_k = 1/[\mathbf{P}_{A^T}^\perp]_{kk}$ and
$\boldsymbol{\epsilon}^s = [\epsilon^s_1 \,\, \epsilon^s_2 \,\, \ldots \,\, \epsilon^s_{N}]^T  = \mathbf{P}_{A^T}^\perp \mbf{x}$.

\textbf{Cross-validation Procedure:} Let $P_{\max} \leq N - 2 $ denote the maximum polynomial order to be tested for SG-filtering. The procedure starts with the $0$th order. For this order, the $\mbf{A}$ matrix is $1 \times N$ dimensional and the residual error vector $\boldsymbol{\epsilon}^s = \mathbf{P}_{A^T}^\perp \mbf{x}$ is formed by projecting the input $\mbf{x}$ onto the $N-1$ dimensional null space of this particular $\mbf{A}$ matrix. The weight $\gamma_k = 1/[\mathbf{P}_{A^T}^\perp]_{kk}$ in \mref{Tpe} is calculated as the reciprocal of the $k$th diagonal entry of the matrix $\mathbf{P}_{A^T}^\perp$. Note that, we have $1/\gamma_k  = [\mathbf{P}_{A^T}^\perp]_{kk} =
\mbf{e}_k^T \mathbf{P}_{A^T}^\perp \mbf{e}_k =
|| \mathbf{P}_{A^T}^\perp \mbf{e}_k ||^2  < || \mathbf{P}_{A^T}^\perp \mbf{e}_k ||^2 + || \mathbf{P}_{A^T} \mbf{e}_k ||^2 = ||\mbf{e}_k ||^2 = 1$; that is $\gamma_k > 1$, since projection onto a sub-space is a contractive operator reducing the norm of the input.
The total squared prediction error $T^p_\epsilon$ for the $0$th order is calculated from \mref{Tpe}. The process is then repeated for higher orders up to the maximum order $P_{\max}$. Once $T^p_\epsilon$ is calculated for all orders, the procedure returns the order with the minimum $T^p_\epsilon$ value as the selected order.

\textbf{Bias-variance trade-off:} The total squared prediction error $T^p_\epsilon$  in \mref{Tpe} depends on $\boldsymbol{\epsilon}^s = \mathbf{P}_{A^T}^\perp \mbf{x}$ and $1/\gamma_k = || \mathbf{P}_{A^T}^\perp \mbf{e}_k||^2$. As the candidate polynomial order $P$ is gradually increased, the number of rows in the matrix $\mbf{A}$ in \mref{SGfiltpregen} increases and this reduces the dimension of its null space (nullity). Since $\mathbf{P}_{A^T}^\perp$ is the projector onto the null space of $\mbf{A}$, both $||\boldsymbol{\epsilon}^s||^2$ and $1/\gamma_k$ decrease as the model order increases. Hence, the factor $\gamma_k$ in $T^p_\epsilon$ increases monotonically with the model order, while the other factor, the squared residual error decreases monotonically in norm. This results in a trade-off that effectively acts as a penalty function, enabling model order selection. The trade-off is illustrated with a numerical experiment in the next section.

\textbf{Computational Complexity:} The conventional $N$-fold CV implementation requires $O(N^5)$ operations. The efficient algorithm in Table~\ref{algSGordwithCV}, based on GS orthogonalization via Householder QR, reduces the complexity to $O(N^3)$, or to $O(N^2)$ when the GS step is precomputed. Precomputing the GS step requires storing an $N \times (N-1)$ matrix. In our tests, the precomputed version yielded 36-fold and 400-fold speedups for $N=10$ and $N=20$, respectively (27-fold and 310-fold when GS-step is computed on the fly). These values are implementation and hardware dependent, but they indicate the substantial efficiency gain of the proposed method. Further details are available in the appendices of the extended version in \cite{SGorderMATLABcode} along with MATLAB codes.

\section{Numerical Results}
\label{numressec}
We compare the proposed method with the Bayesian Information Criterion (BIC), which is the state-of-the-art model order selection rule for the polynomial fitting problem \cite{StoicaOnProper2021}. BIC approximates the model likelihood by treating the model parameters as random variables and applies either high SNR or large sample size asymptotic theory to develop practical model order estimators \cite{ModelOrderSPM2004}. In our performance comparisons, we include both large sample size variant \BICN,  and the low noise variance (high SNR) variant \BICSNR, as given in \cite{StoicaOnProper2021}:
\be
\BICN(P) & = & (P+1)^2\ln(N) + N\ln\left(\frac{|| \boldsymbol{\epsilon}^s ||^2}{N}\right)  \nn \\
\BICSNR(P) & = & N \ln\left(\frac{|| \boldsymbol{\epsilon}^s ||^2}{N}\right)
+ \max(0,-(P+2)\ln\left(\frac{|| \boldsymbol{\epsilon}^s ||^2}{N}\right)) \nn
\ee
Here $P$ denotes the polynomial order, $N$ is the processing window length, and $\boldsymbol{\epsilon}^s = \mathbf{P}_{A^T}^\perp \mbf{x}$ is the residual error vector after smoothing, as described in Section~\ref{SecMain}.

In the experiment conducted, the samples $y[n]$ are noisy observations of the cubic function $x(t) = 0.01 t^3 + 1$. The function is sampled with the sampling period of $T_s=1$, i.e.,  $x[n] = x(nT_s)$ and the observations $y[n]$ are collected under white Gaussian observation noise $w[n]$, i.e., $y[n] = x[n] + w[n]$. The noise $w[n]$ is zero mean with variance $\sigma_w^2$. We assume that $y[n]$ samples are obtained over the interval $[-N/2, N/2-1]$ for even $N$ and over the interval $[-(N-1)/2,(N-1)/2]$ for odd $N$. The candidate model orders include all polynomial degrees from $0$ to $N-2$.
\ifmyshow \textcolor{red}{--- Figures 2 and 3 about here --- } \fi

\begin{figure}[t]
\centering
\includegraphics[height=10cm]{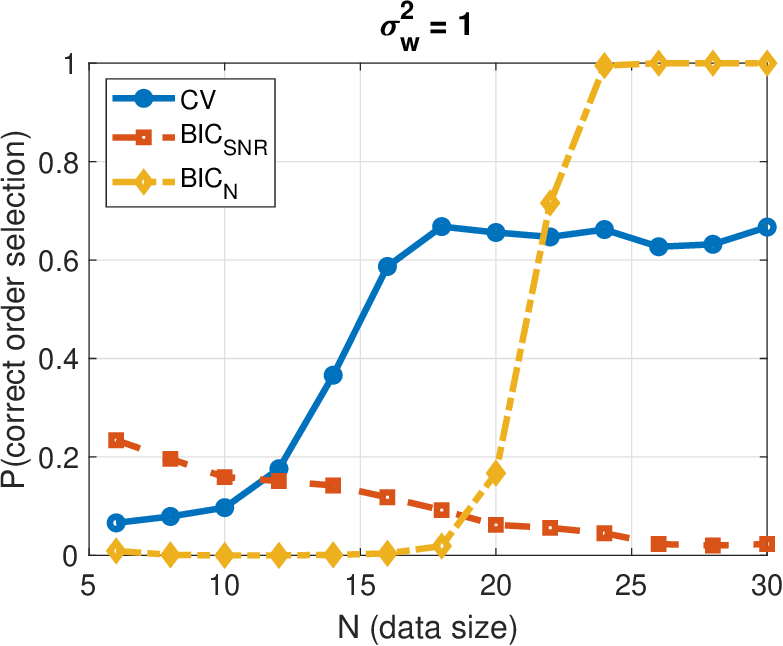}
\caption{Probability of correct model order detection at fixed noise variance as the sample size (processing window length) $N$ increases. The suggested method outperforms the asymptotically optimal $\BICN$ at intermediate, that is non-asymptotic, values of $N$.}
\label{fignumresN}
\end{figure}

\begin{figure}[t]
\centering
\includegraphics[height=10cm]{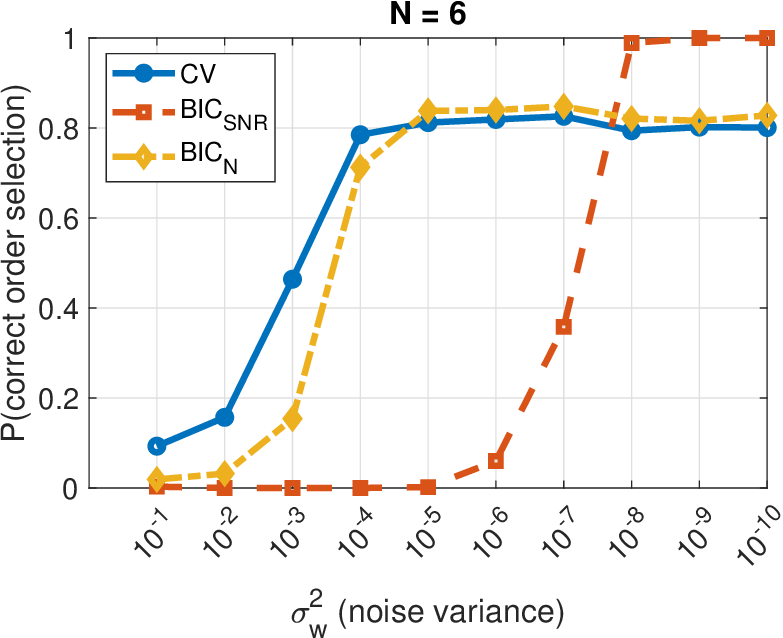}
\caption{Probability of correct model order detection at a fixed sample size of $N=6$ as the noise variance decreases. The suggested method outperforms the asymptotically optimal $\BICSNR$ at intermediate, that is non-asymptotic, SNR values.}
\label{fignumresSNR}
\end{figure}

Figure~\ref{fignumresN} shows the probability of correctly detecting model order for a fixed noise variance of $1$ as the processing window size $N$ increases. We observe that cross-validation (CV) method outperforms both $\BICN$ and $\BICSNR$ for intermediate values of $N$. For asymptotically large $N$ values, $\BICN$ outperforms all others as expected; since it is designed to be a consistent model estimator in the large sample size regime \cite{StoicaOnProper2021,InconsMDLKay2011}. Therefore, performance superiority over $\BICN$ is not expected at high sample sizes. Figure~\ref{fignumresSNR} shows the same comparison when the sample size $N$ is fixed at $6$ samples, while the noise variance is gradually reduced from $10^{-1}$ to $10^{-10}$. As expected, $\BICSNR$ outperforms the other methods at sufficiently low noise variance levels due to its asymptotic optimality in the high-SNR regime. We observe that the proposed cross-validation based order detector outperforms both $\BICN$ and $\BICSNR$ at moderate sample sizes and noise variance levels, a significant result for the applications limited to non-asymptotic regimes.

\begin{figure}[h]
\centering
\includegraphics[height=10cm]{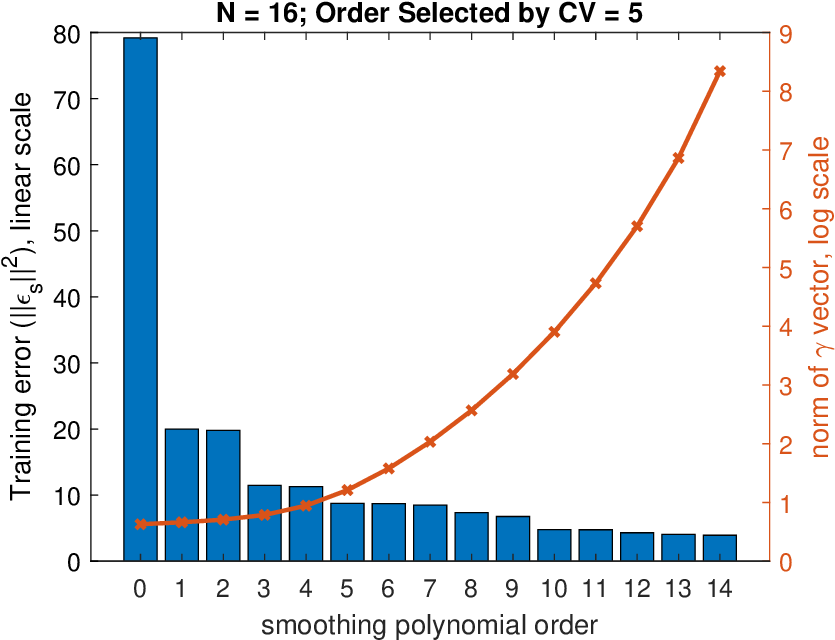}
\caption{An illustration for the bias-variance trade-off: Smoothing error decreases with the polynomial order while associated weights increases. This results in a trade-off point between a perfect match to the training data (small bias, large variance) and a preference of simpler models with low ordered polynomials (large bias, small variance).}
\label{figbias-var}
\end{figure}

Figure~\ref{figbias-var} illustrates the bias-variance tradeoff. Here, the results for a single Monte-Carlo run with the data length of $N=16$ are shown. We observe that training error gets smaller with the smoothing polynomial order; while the associated weight for higher orders increases. The trade-off between two enables the selection of medium ordered polynomials, which is a 5th degree polynomial in this case. \ifmyshow \textcolor{red}{--- Figure 4 about here --- } \fi

\section{Summary and Conclusions}
A simple model order (polynomial order) selection procedure with $N$-fold cross-validation is derived for SG filtering. The expression is based on the minimum norm formulation of SG filters and can be computed efficiently. An order-recursive implementation of the proposed technique, utilizing Gram-Schmidt orthogonalization, is given in Table~\ref{algSGordwithCV}. A ready-to-use implementation is also given in an online code repository \cite{SGorderMATLABcode}. The absence of tuning parameters and the strong performance in the non-asymptotic regimes makes the proposed method attractive.

From technical viewpoint, the performance of BIC-based order estimators is known to degrade when either $N$ (sample size) or signal-to-noise-ratio (SNR) is not sufficiently high. This limitation arises from the asymptotic approximations used in model likelihood calculations. The numerical results demonstrate that in regimes where BIC-based estimators falter, namely moderate $N$ or moderate SNR, the proposed cross-validation technique can be adopted for improved performance. Although cross-validation is often regarded as an empirical method with loose theoretical grounding, it is interesting to note that its application to the SG filtering problem forms a rare case in which the method admits a theoretically justifiable trade-off explanation for its good empirical performance.

For future work, the robustness of CV method, say to impulsive noise (see \cite{SGorderMATLABcode} for initial findings); the case of irregularly sampled and the extensions to other SG-filters can be studied.
\ifmyshow \textcolor{red}{--- Table 2 about here --- } \fi

\begin{table}[!t]\centering
\caption{Model order selection for SG-filtering via cross-validation. See \cite{SGorderMATLABcode} for MATLAB code.} \label{algSGordwithCV}{
\begin{algorithm}[H]
\DontPrintSemicolon\SetCommentSty{}
\SetKwInOut{Input}{Input}
\SetKwInOut{Output}{Output}
\SetKwInOut{Return}{Return}
{\small\Input{$\mathbf{x}=[x_1,x_2,\ldots,x_N]^T$, $P_{\max}$, $\mbf{Q} = [\mathbf{q}_0, \mathbf{q}_1, \ldots,  \mathbf{q}_{P_{\max}}]$ (Optional)}
	\Output{$\textrm{best-order}$}
	
     \tcp{Gram-Schmidt orthogonalization (if needed)}
     \If{$\mbf{Q}$ is not available}
                {$[Q,R] = \textrm{qr}(A^T)$ \tcp*{QR decomposition of $\mathbf{A}^T$, see \mref{SGfiltgen}}}

    $\hat{\mbf{x}}=$ zeros($N$,1); $\mathbf{d}=$ zeros($N$,1) \tcp*{Initialization}
	\For(\tcp*[f]{Recursion}){$k\leftarrow 0$ \KwTo $P_{\max}$}{
		$\hat{\mbf{x}} = \hat{\mbf{x}} + \mbf{q}_k\mbf{q}_k^T\mbf{x}$
        \tcp*{$\hat{\mbf{x}}$ : Projection of $\mbf{x}$ to $(k+1)$-dim subspace}

        $ \boldsymbol{\epsilon}^s = \mbf{x} - \hat{\mbf{x}} $
        \tcp*{Smoothing error}

        $ \mathbf{d}  = \mathbf{d} + \mbf{q}_k.*\mbf{q}_k$
        \tcp*{$\mbf{d} = \textrm{diag}(\mathbf{P}_{A^T})$}

        $ \boldsymbol{\epsilon}^p =  \boldsymbol{\epsilon}^s./ ( 1 -  \mbf{d})$
         \tcp*{.* and ./ : element-wise multiplication and division, see \mref{SGperr1}}
		
		$T^p_\epsilon(k) = || \boldsymbol{\epsilon}^p ||^2$
		\tcp*{see \mref{Tpe}}}

	\Return{$\textrm{best-order} = \arg\!\min_{0\leq k \leq P_{\max}} T^p_\epsilon(k) $}
}	
\end{algorithm}}
\end{table}

%\newpage
{\small
%\bibliography{SG_filter2-cross-validation}

\begin{thebibliography}{10}
\expandafter\ifx\csname url\endcsname\relax
  \def\url#1{\texttt{#1}}\fi
\expandafter\ifx\csname urlprefix\endcsname\relax\def\urlprefix{URL }\fi
\expandafter\ifx\csname href\endcsname\relax
  \def\href#1#2{#2} \def\path#1{#1}\fi

\bibitem{golay64}
A.~Savitzky, M.~J.~E. Golay, {Smoothing and Differentiation of Data by
  Simplified Least Squares Procedures}, Analytical Chemistry 36~(8) (1964)
  1627--1639.

\bibitem{OrfaBOOK}
S.~J. Orfanidis, {Introduction to Signal Processing}, Prentice Hall, 1995.

\bibitem{ccandanSG}
C.~Candan, H.~Inan, {A unified framework for derivation and implementation of
  Savitzky-Golay filters}, Signal Processing 104 (2014) 203--211.

\bibitem{Schafer2011}
R.~Schafer, {What Is a Savitzky-Golay Filter? [Lecture Notes]}, {IEEE} Signal
  Process. Mag. 28~(4) (2011) 111--117.
\newblock \href {https://doi.org/10.1109/MSP.2011.941097}
  {\path{doi:10.1109/MSP.2011.941097}}.

\bibitem{SGappPLL2022}
K.~Hasan, S.~T. Meraj, M.~M. Othman, M.~S.~H. Lipu, M.~A. Hannan, K.~M.
  Muttaqi, {Savitzky-Golay Filter-Based PLL: Modeling and Performance
  Validation}, IEEE Trans. Instrumentation and Measurement 71 (2022) 1--6.
\newblock \href {https://doi.org/10.1109/TIM.2022.3196946}
  {\path{doi:10.1109/TIM.2022.3196946}}.

\bibitem{AdapSG2021}
A.~John, J.~Sadasivan, C.~S. Seelamantula, {Adaptive Savitzky-Golay Filtering
  in Non-Gaussian Noise}, {IEEE} Trans. Signal Process. 69 (2021) 5021--5036.

\bibitem{GStrang2003}
P.-O. Persson, G.~Strang, {Smoothing by Savitzky-Golay and Legendre Filters},
  in: Mathematical Systems Theory in Biology, Comm., Computation, Finance,
  Springer, NY, 2003, pp. 301--315.

\bibitem{WindowSelect2020}
M.~Sadeghi, F.~Behnia, R.~Amiri, {Window Selection of the Savitzky-Golay
  Filters for Signal Recovery From Noisy Measurements}, IEEE Trans. Instrum.
  Meas. 69~(8) (2020) 5418--5427.

\bibitem{SG_IEEESP2013}
S.~Krishnan, C.~Seelamantula, {On the Selection of Optimum Savitzky-Golay
  Filters}, {IEEE} Trans. Signal Process. 61~(2) (2013) 380--391.
\newblock \href {https://doi.org/10.1109/TSP.2012.2225055}
  {\path{doi:10.1109/TSP.2012.2225055}}.

\bibitem{Barberbook}
D.~Barber, {Bayesian Reasoning and Machine Learning}, Cambridge Univ. Press,
  2012.

\bibitem{SGorderMATLABcode}
C.~Candan, \href{https://arxiv.com}{{Polynomial Order Selection for
  {Savitzky-Golay} Smoothers via N-fold Cross-Validation (Extended Version)}},
  {MATLAB} source codes: \url{https://doi.org/10.24433/CO.1732394.v2} (2025).
\newline\urlprefix\url{https://arxiv.com}

\bibitem{StoicaOnProper2021}
P.~Stoica, P.~Babu, {On the Proper Forms of BIC for Model Order Selection},
  {IEEE} Trans. Signal Process. 60~(9) (2012) 4956--4961.
\newblock \href {https://doi.org/10.1109/TSP.2012.2203128}
  {\path{doi:10.1109/TSP.2012.2203128}}.

\bibitem{ModelOrderSPM2004}
P.~Stoica, Y.~Selen, Model-order selection: a review of information criterion
  rules, {IEEE} Signal Process. Mag. 21~(4) (2004) 36--47.
\newblock \href {https://doi.org/10.1109/MSP.2004.1311138}
  {\path{doi:10.1109/MSP.2004.1311138}}.

\bibitem{InconsMDLKay2011}
Q.~Ding, S.~Kay, {Inconsistency of the MDL: On the Performance of Model Order
  Selection Criteria With Increasing Signal-to-Noise Ratio}, {IEEE} Trans.
  Signal Process. 59~(5) (2011) 1959--1969.

\end{thebibliography}
%\biboptions{sort&compress}

}

\appendix

\section{Additional Notes on Efficient Implementation and its Complexity}
The Gram-Schmidt (GS) orthogonalization procedure is known to be numerically unstable. We use its Householder QR decomposition implementation which is a stable and widely preferred implementation adopted by many numerical packages. For example, the qr.m function in MATLAB uses LAPACK (a free and open-source numerical library).

\textbf{On Pivoting:} In our application, we use QR decomposition without pivoting, since the columns of $\mbf{A}^T$ in \mref{SGfiltgen} must remain in increasing polynomial order. This ordering ensures that the associated projection spaces are nested; that is, the projection space, say, for the third order polynomials contains the space for the second order polynomials.

Table~\ref{comp_CO} presents computation time comparisons for different implementations of the proposed method. The efficient implementation in Table~\ref{algSGordwithCV} has two variations, one with GS orthogonalization step and one without, in relation to the if-statement in step~1 of Table~\ref{algSGordwithCV}. To assess the efficiency of the suggested implementation, we executed the same code on two different environments (a standard PC with i7 CPU and a high-performance cloud server) and recorded the runtimes. All runtimes in Table~\ref{comp_CO} are given in milliseconds.

The improvement factors, given in brackets, show the ratio of the runtimes for the conventional and efficient implementation. In both environments, we observe a substantial improvement in computational efficiency with the proposed efficient implementation. As a cautionary note, the exact improvement factor depends strongly on the software, hardware and operating system used. Table~\ref{comp_CO} is intended to assist practitioners by providing a realistic expectation of the performance improvement. (In the main text, we report the smaller of the two observed improvement factors to remain conservative.) The computational time comparisons in Table~\ref{comp_CO} can be repeated by running the code in the CodeOcean capsule \cite{SGorderMATLABcode} or in any other environment.

\begin{table}[]
\caption{Computation time comparisons for the proposed method:  Efficient implementation proposed in Table~\ref{algSGordwithCV} vs. conventional implementation}
\centering
\begin{tabular}{llll}
\cline{2-4}
\multicolumn{1}{l|}{}               & \multicolumn{3}{c|}{\textbf{on codeocean cloud server under Linux, Matlab 2023b}}                                                                                                                                                                                                            \\ \hline
\multicolumn{1}{|l|}{}              & \multicolumn{1}{c|}{\textbf{\begin{tabular}[c]{@{}c@{}}Conventional\\ (msec)\end{tabular}}} & \multicolumn{1}{c|}{\textbf{\begin{tabular}[c]{@{}c@{}}Proposed\\ without GS\\ Step\\ (msec)\end{tabular}}} & \multicolumn{1}{c|}{\textbf{\begin{tabular}[c]{@{}c@{}}Proposed\\ with GS\\ Step\\ (msec)\end{tabular}}} \\ \hline
\multicolumn{1}{|l|}{\textbf{N=5}}  & \multicolumn{1}{l|}{0.069389}                                                               & \multicolumn{1}{l|}{0.007356 ($\sim$9 folds)}                                                               & \multicolumn{1}{l|}{0.011481 ($\sim$6 folds)}                                                            \\ \hline
\multicolumn{1}{|l|}{\textbf{N=10}} & \multicolumn{1}{l|}{0.44692}                                                                & \multicolumn{1}{l|}{0.00703 ($\sim$63 folds)}                                                               & \multicolumn{1}{l|}{0.012455 ($\sim$35 folds)}                                                           \\ \hline
\multicolumn{1}{|l|}{\textbf{N=15}} & \multicolumn{1}{l|}{4.9698}                                                                 & \multicolumn{1}{l|}{0.007184 ($\sim$691 folds)}                                                             & \multicolumn{1}{l|}{0.015694 ($\sim$316 folds)}                                                          \\ \hline
\multicolumn{1}{|l|}{\textbf{N=20}} & \multicolumn{1}{l|}{15.72}                                                                  & \multicolumn{1}{l|}{0.008934 ($\sim$1759 folds)}                                                            & \multicolumn{1}{l|}{0.020421 ($\sim$769 folds)}                                                          \\ \hline
                                    &                                                                                             &                                                                                                             &                                                                                                          \\ \cline{2-4}
\multicolumn{1}{l|}{}               & \multicolumn{3}{c|}{\textbf{on i7 PC under Windows OS, Matlab 2015b}}                                                                                                                                                                                                                        \\ \hline
\multicolumn{1}{|l|}{\textbf{}}     & \multicolumn{1}{c|}{\textbf{\begin{tabular}[c]{@{}c@{}}Conventional\\ (msec)\end{tabular}}} & \multicolumn{1}{c|}{\textbf{\begin{tabular}[c]{@{}c@{}}Proposed\\ without GS\\ Step\\ (msec)\end{tabular}}} & \multicolumn{1}{c|}{\textbf{\begin{tabular}[c]{@{}c@{}}Proposed\\ with GS\\ Step\\ (msec)\end{tabular}}} \\ \hline
\multicolumn{1}{|l|}{\textbf{N=5}}  & \multicolumn{1}{l|}{0.24389}                                                                & \multicolumn{1}{l|}{0.019663 ($\sim$12 folds)}                                                              & \multicolumn{1}{l|}{0.027268 ($\sim$8 folds)}                                                            \\ \hline
\multicolumn{1}{|l|}{\textbf{N=10}} & \multicolumn{1}{l|}{1.0216}                                                                 & \multicolumn{1}{l|}{0.027939 ($\sim$36 folds)}                                                              & \multicolumn{1}{l|}{0.036634 ($\sim$27 folds)}                                                           \\ \hline
\multicolumn{1}{|l|}{\textbf{N=15}} & \multicolumn{1}{l|}{7.3221}                                                                 & \multicolumn{1}{l|}{0.040924 ($\sim$178 folds)}                                                             & \multicolumn{1}{l|}{0.047323 ($\sim$154 folds)}                                                          \\ \hline
\multicolumn{1}{|l|}{\textbf{N=20}} & \multicolumn{1}{l|}{22.081}                                                                 & \multicolumn{1}{l|}{0.055324 ($\sim$400 folds)}                                                             & \multicolumn{1}{l|}{0.071133 ($\sim$310 folds)}                                                          \\ \hline
\end{tabular}
\label{comp_CO}
\end{table}

\textbf{On Storage Requirements:} The efficient implementation without the GS-step requires
additional storage for an $N\times (P+1)$ dimensional matrix which corresponds to the $\mbf{Q}$ matrix obtained from the QR decomposition of $\mbf{A}^T$ matrix. Here, $P$ is the maximum polynomial order to be tested. We have $P \leq N-2$, since $(N-2)$'th degree polynomials have $N-1$ degrees of freedom and we use $N-1$ samples to predict the remaining single sample in the validation set. Therefore, the storage requirement of the efficient scheme, when using a pre-computed $\mbf{Q}$ matrix, is at most an $N \times (N-1)$ dimensional matrix in the worst case.

\textbf{Beyond N-fold CV operation:} This study examines the CV setting in which only one sample is placed in the validation set and the remaining $N-1$ samples are used to predict that sample. Naturally, this framework can be extended to the cases with multiple samples in the validation set.

If $k$ samples are placed in the validation set; then the remaining $N-k$ samples are used for their prediction and the total number of possible validation/training set partitions is ${N\choose k}$. Therefore, extending from $k=1$ to $k=2$ requires repeating the prediction operation ${N\choose 2} = N(N-1)/2$ times, instead of $N$ times, highlighting the necessity of an efficient implementation.

An efficient implementation for the general scheme can be given by updating the $\mbf{A}$ matrix in \mref{SGfiltpregen} by introducing $k$ columns with all zeros instead of a single column. With this modification to $\mathbf{A}$, the matrix inversion lemma can be applied $k$ times (instead of once) to obtain an efficient implementation, generalizing the approach proposed in this work.

However, for our SG-smoothing application considered here, the N-fold CV procedure which uses the maximum number of samples in the training set while incurring the least computational overhead, appears to be the natural choice for model order detection. Therefore, this extension may be viewed primarily as an academic generalization for completeness, with limited practical engineering relevance.

\section{On Algorithm Robustness}
We present preliminary findings on the robustness of the proposed model order selection algorithm. The numerical experiment is identical to the one described in Section~\ref{numressec}.

As in Section~\ref{numressec}, the samples $y[n]$ are noisy observations of the cubic function $x(t) = 0.01 t^3 + 1$. The function is sampled with a sampling period of $T_s=1$, i.e.,  $x[n] = x(nT_s)$ and noisy observations $y[n]$ are then obtained. The noise is independent and identically distributed; but, different from the experiment in Section~\ref{numressec}, the noise $W$ is modeled as a mixture of two zero mean Gaussian random variables with variances $\sigma_w^2$ and $\sigma_i^2$:
\be
f_W(w) = (1-p_i)\times N(w;0, \sigma_w^2) + p_i \times N(w;0, \sigma_i^2). \nn
\ee
Here $N(w;\mu, \sigma^2)$ denotes to the Gaussian distribution with mean $\mu$ and variance $\sigma^2$. We assume a latent Bernoulli random variable $B_i \in \{0,1\}$ with $P(B_i = 1) = p_i$. The random variable is independent of all other random variables and serves as an indicator variable for impulsive noise. That is, when $B_i = 1$ (impulsive noise) and $B_i=0$ (nominal noise), the random variable $W$ has the density of $f_{W|B_i=1} (w | B_i = 1) =  N(w;0, \sigma_i^2)$ (impulsive  noise distribution) and $f_{W|B_i=0} (w | B_i = 0) =  N(w;0, \sigma_w^2)$ (nominal noise distribution), respectively. We assume that $\sigma_i^2 \gg \sigma_w^2$.

\begin{figure}[t]
\centering
\includegraphics[height=9cm]{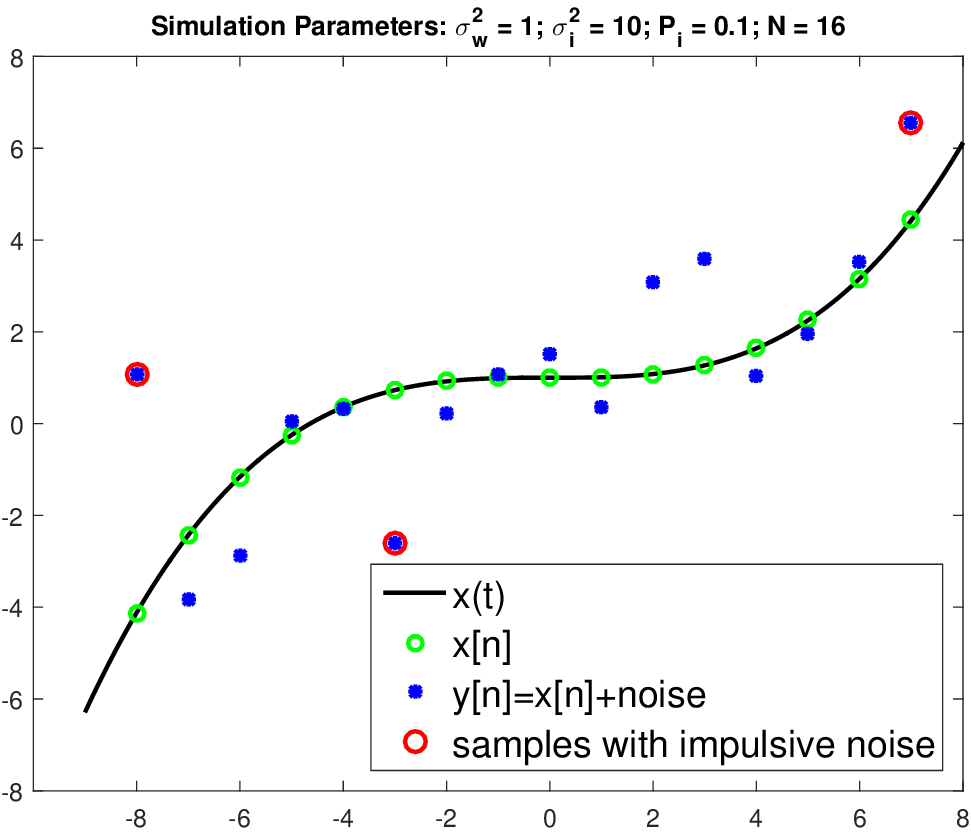}
\caption{A 3rd order polynomial, its noiseless/noisy samples. To test algorithm robustness, noise is selected as a mixture of two Gaussian distributions. One of mixture components represents the nominal noise with variance $\sigma_w^2$ and the other one represents impulsive noise with much higher variance $\sigma_i^2$. The impulsive noise can occur independently with probability $p_i$ at every sample.}
\label{figimpnoise1}
\end{figure}

Figure~\ref{figimpnoise1} shows $x(t)$, its samples $x[n]$ and the noisy samples $y[n]$ corrupted by the mixture noise. The samples $y[n]$ marked with red circles indicate those contaminated by impulsive noise. In this figure, the impulsive noise has 10 times higher variance than nominal noise variance $\sigma_w^2=1$. The probability of an impulsive noise occurrence is $p_i = 0.1$.

\begin{figure}[h]
\vspace*{-1cm}
    \centering
    \begin{subfigure}{
    \includegraphics[height=7cm]{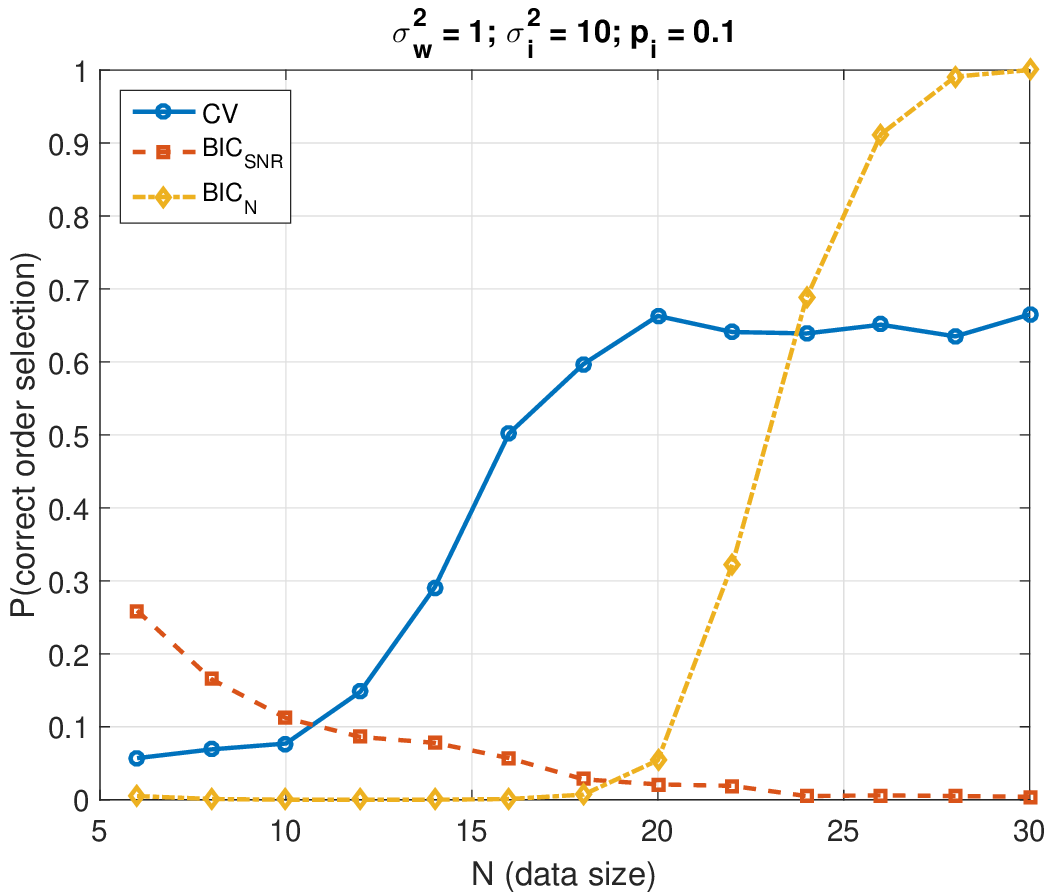}}
    \end{subfigure}
    \\
    \begin{subfigure}{
    \includegraphics[height=7cm]{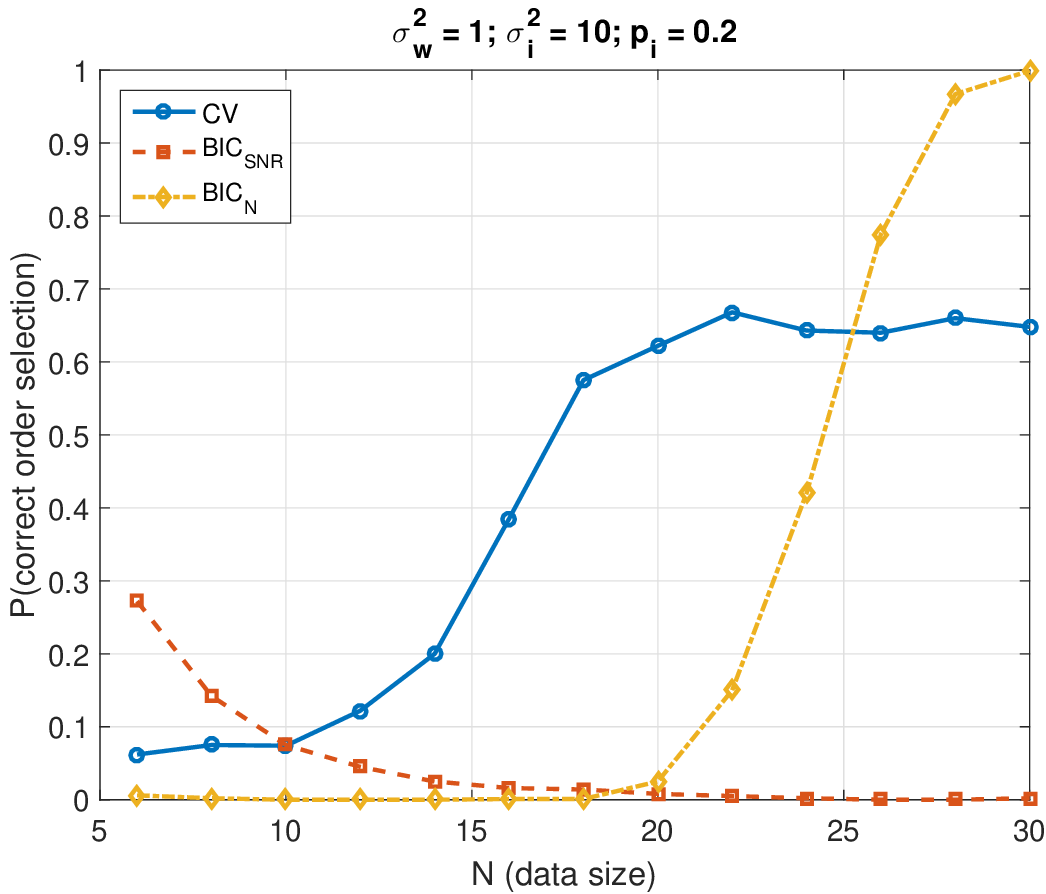}}
    \end{subfigure}
    \\
    \begin{subfigure}{
    \includegraphics[height=7cm]{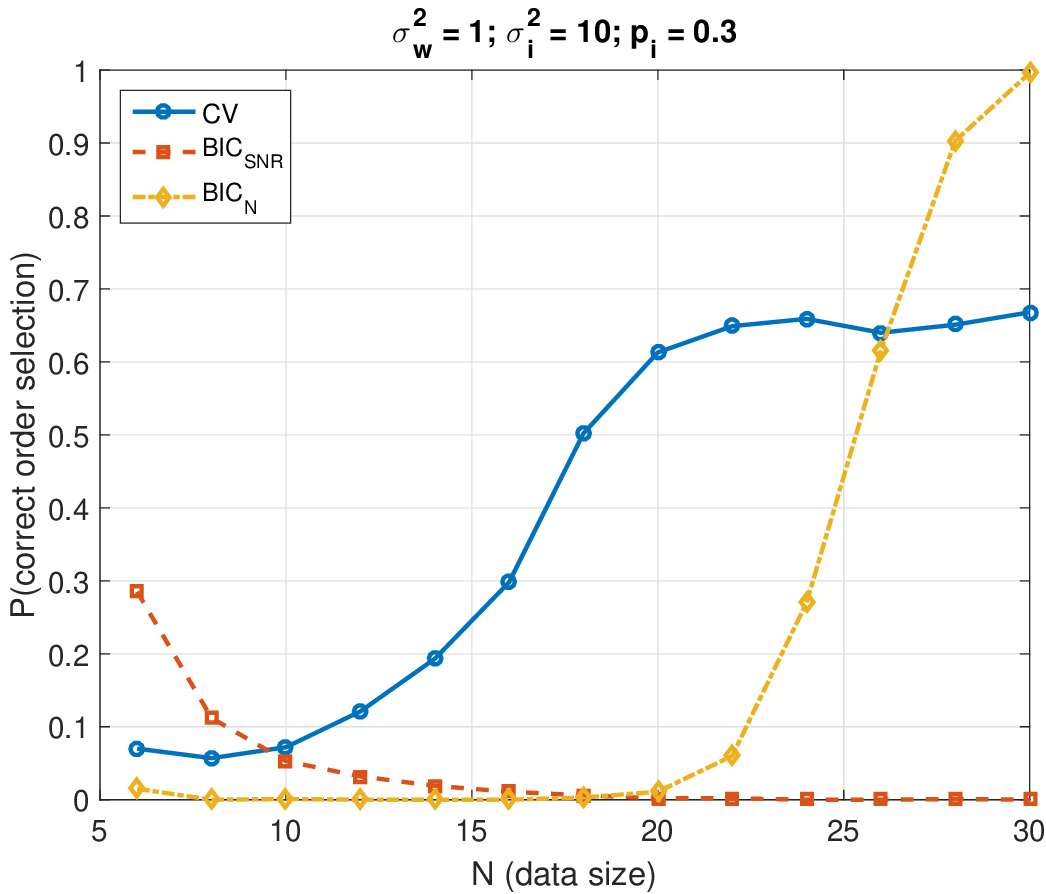}}
    \end{subfigure}
\caption{Probability of correct model order detection vs. the window size ($N$) at different impulsive noise occurrence probability values ($p_i$).}
\label{figimpnoise2}
\end{figure}

Figure~\ref{figimpnoise2} shows the probability of correct model order detection as the window size $N$ increases, for different $p_i$ values. As expected, $\BICSNR$ performs poorly because the sample SNR is fixed across all three simulations shown in Figure~\ref{figimpnoise2}. $\BICN$ performs well in the asymptotic regime of large $N$; however, as the model mismatch increases (the impulsive noise probability increases), the threshold value of $N$ required to achieve good performance also increases. In contrast, the proposed method has nearly identical performance across all cases in Figure~\ref{figimpnoise2}.

Our initial explanation for the observed robustness of the CV method is that samples contaminated by impulsive noise can not be predicted reliably unless the model order is extremely high. Hence, for medium values of $N$ (data window size), samples affected by impulsive noise suffer from large prediction errors regardless of the value of $N$. In N-fold CV procedure, some samples are well predicted (samples contaminated by nominal noise); while others are poorly predicted (samples contaminated by impulsive noise). As $N$ increases, the well-predicted samples become even more accurately predicted, whereas the impulsive noise samples remain poorly predicted. This leads to the observed robustness or insensitivity of the CV method to impulsive noise.

\begin{figure}[h]
\vspace*{-1cm}
    \centering
    \begin{subfigure}[weak \& infrequent impulsive noise]{
    \includegraphics[height=6.75cm]{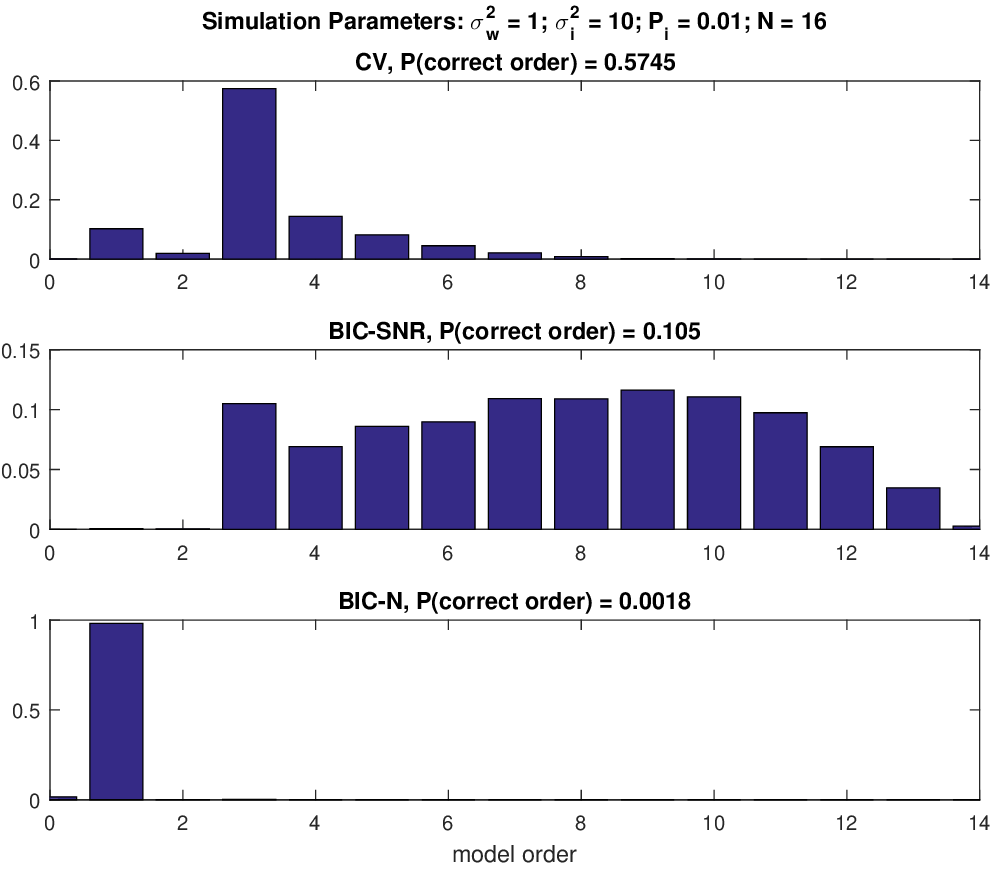}}
    \end{subfigure}
    %\hspace(1cm}
    \begin{subfigure}[strong \& infrequent impulsive noise]{
    \includegraphics[height=6.75cm]{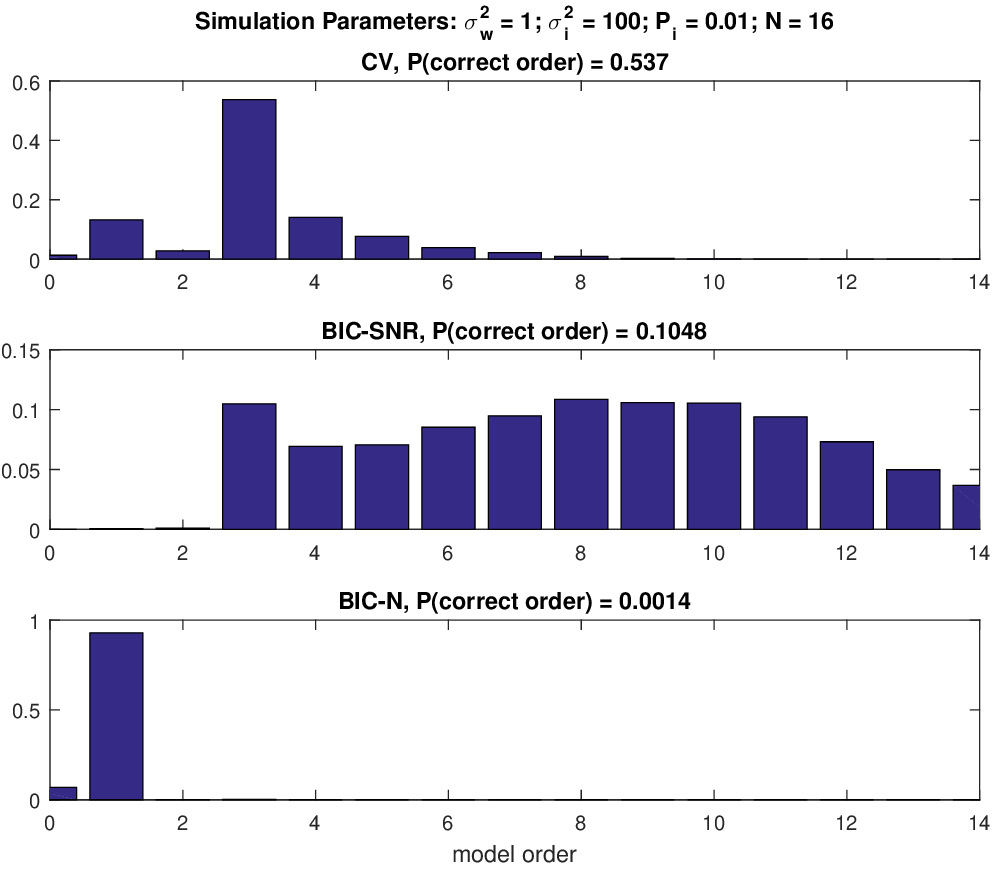}}
    \end{subfigure}
    \\
    \begin{subfigure}[weak \& frequent impulsive noise]{
    \includegraphics[height=6.75cm]{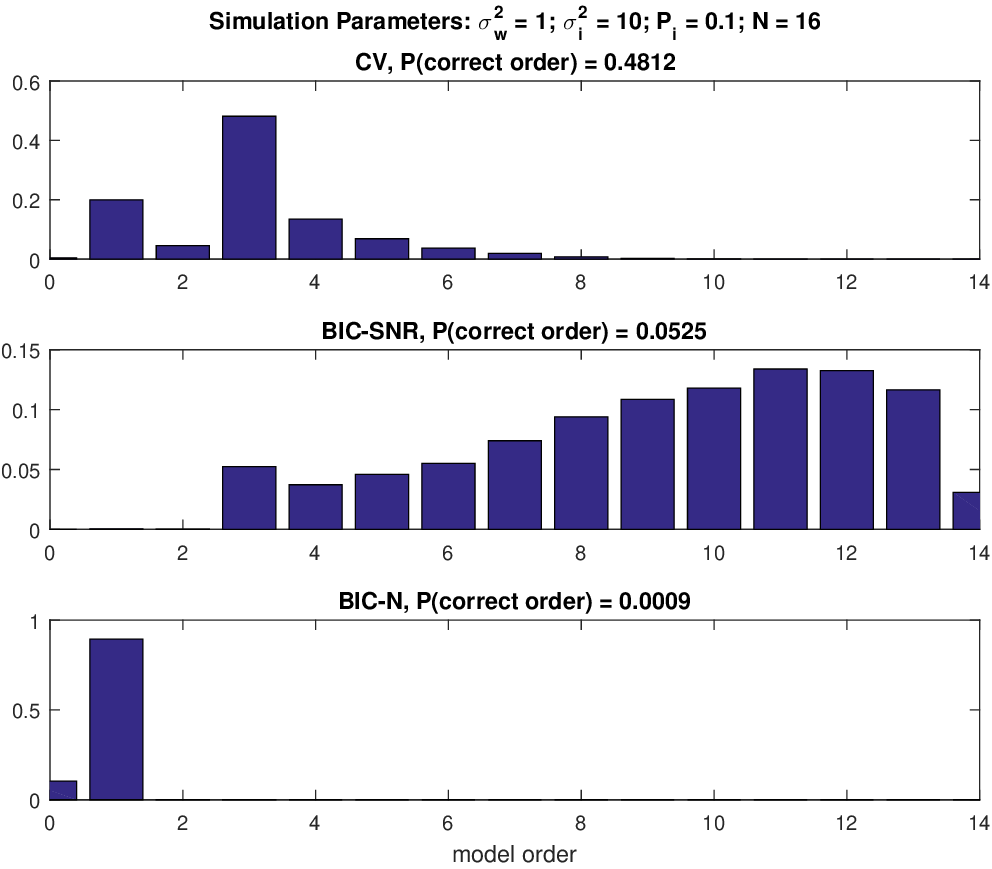}}
    \end{subfigure}
    %\hspace(1cm}
    \begin{subfigure}[strong \& frequent impulsive noise]{
    \includegraphics[height=6.75cm]{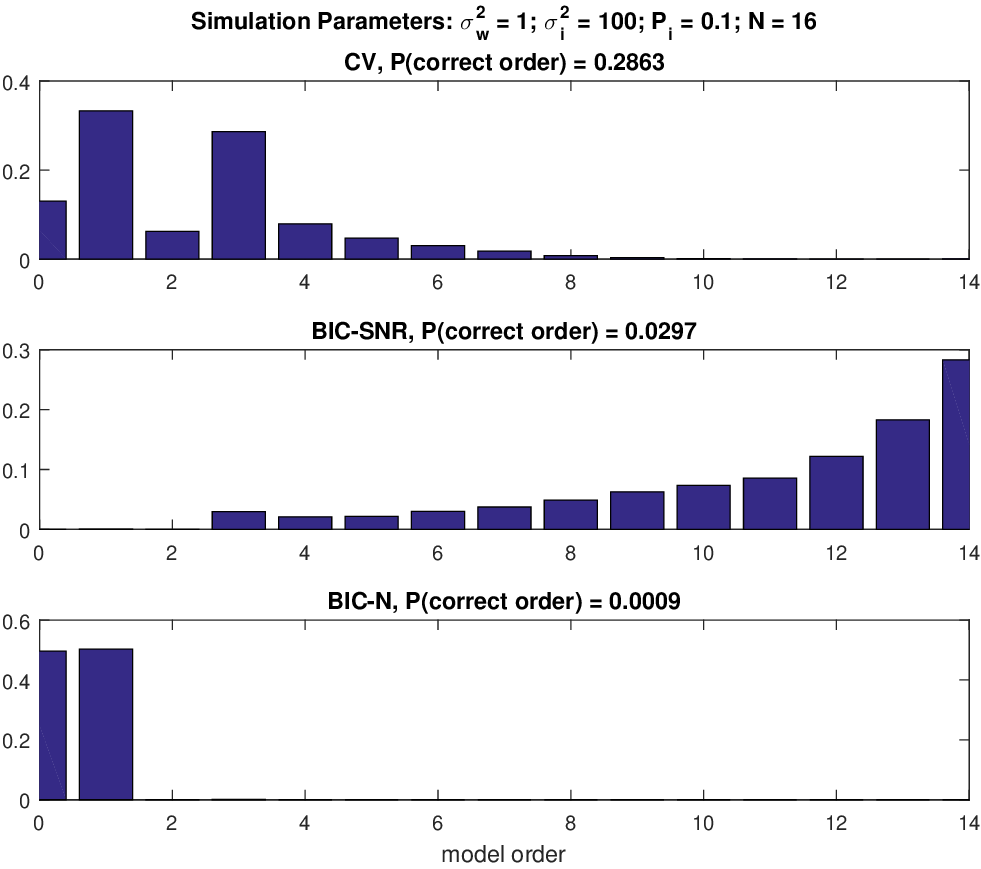}}
    \end{subfigure}
\caption{Distribution of detected orders by proposed and other methods for weak/strong impulsive noise power ($\sigma_i^2 \in \{10,100\}$) and two different probability values for impulsive noise occurrence ($p_i \in \{0.01, 0.1\}$).}
\label{figimpnoise3}
\end{figure}

Figure~\ref{figimpnoise3} examines the effect of impulsive noise power and occurrence frequency at a fixed window length of $N=16$. The impulsive noise variance power is set to either $10$ or $100$ (i.e., $\sigma_i^2 \in \{10, 100\}$) to represent weak and strong impulsive noise conditions. The impulsive noise occurrence probability is set to either $0.01$ or $0.1$ to reflect infrequent and frequent impulsive noise scenarios, respectively.

From Figure~\ref{figimpnoise3}, we observe that the proposed method offers much better robustness than the other methods considered. The $\BICN$ estimator, which is known to exhibit a threshold effect (see Figure~\ref{figimpnoise2}), performs poorly in all cases, as $N=16$ is below its reliable operation threshold. The $\BICSNR$ estimator also performs poorly. It produces an almost uniform distribution of detected model orders under infrequent impulsive noise (cases (a) and (b)) and it tends to select higher than true model order under frequent impulsive noise (cases (c) and (d)) as reflected by the skew of the distribution towards right.

We present these initial robustness findings to assist and encourage further research on this topic.

\end{document}